\pdfoutput=1
\documentclass[11pt]{article}

\usepackage[]{acl}

\usepackage{times}
\usepackage{latexsym}

\usepackage[T1]{fontenc}
\usepackage[utf8]{inputenc}

\usepackage{microtype}

\usepackage{xcolor}
\usepackage{graphicx}
\usepackage{amssymb}
\usepackage{amsmath}
\usepackage{booktabs} % for professional tables
\usepackage{makecell}
\usepackage{multirow}
\usepackage{natbib}
\title{Scaling Properties of Speech Language Models}

\author{Santiago Cuervo \and Ricard Marxer \\
        Universit\'e de Toulon, Aix Marseille Universit\'e, CNRS, LIS, France\\
        \texttt{\{santiago.cuervo, ricard.marxer\}@lis-lab.fr}}
\begin{document}
\maketitle
\begin{abstract}
Speech Language Models (SLMs) aim to learn language from raw audio, without textual resources. Despite significant advances, our current models exhibit weak syntax and semantic abilities. However, if the scaling properties of neural language models hold for the speech modality, these abilities will improve as the amount of compute used for training increases. In this paper, we use models of this scaling behavior to estimate the scale at which our current methods will yield a SLM with the English proficiency of text-based Large Language Models (LLMs). We establish a strong correlation between pre-training loss and downstream syntactic and semantic performance in SLMs and LLMs, which results in predictable scaling of linguistic performance. We show that the linguistic performance of SLMs scales up to three orders of magnitude more slowly than that of text-based LLMs. Additionally, we study the benefits of synthetic data designed to boost semantic understanding and the effects of coarser speech tokenization.
\end{abstract}

\section{Introduction}

\begin{figure}[ht]
\vskip 0.2in
\begin{center}
\includegraphics[width=0.9\columnwidth]{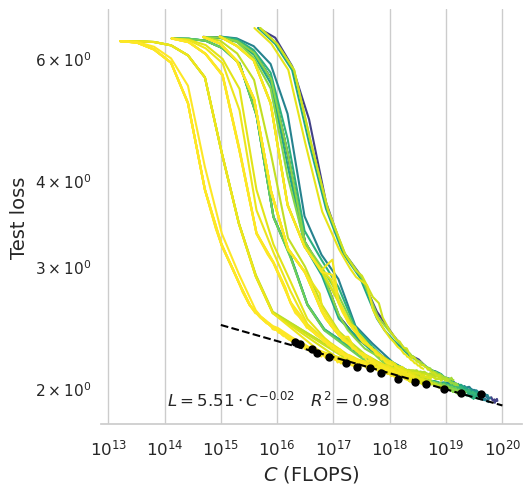}
\caption{Speech Language Models test loss curves for all our single-epoch runs. Axes are in logarithmic scale. The envelope of minimal loss per FLOP (black dots) follows a power law (dashed line).}
\label{fig:test-scale}
\end{center}
\vskip -0.2in
\end{figure}

Inspired by the remarkable ability of preschool children to learn language from raw sensory inputs, \citet{lakhotia-etal-2021-generative} introduced in their seminal paper the \textit{textless NLP} (Natural Language Processing) project. The project aimed to leverage advances in self-supervised speech representation learning for unsupervised unit discovery \cite{hsu2021hubert, w2v-bert} and generative neural language models \cite{brown2020language} to jointly learn the acoustic and linguistic characteristics of a language from audio alone, without access to textual supervision (e.g. lexicon or transcriptions). They formalized this goal in the task of \textit{Generative Spoken Language Modeling} (GSLM), in which a language model is trained on sequences of self-supervised learned speech units.

Despite a significant body of research on these speech-based language models (SLMs) \cite{lakhotia-etal-2021-generative, kharitonov-etal-2022-text, borsos-audiolm, hassid2023textually}, they are still far from matching the syntactic and semantic abilities of text-based systems \cite{hassid2023textually}. Therefore, the promise of textless NLP is yet to be realized. However, if the scaling behavior of text-based neural language models \cite{brown2020language, kaplan2020} holds for the speech modality, we can reasonably expect those abilities to improve as the amount of compute used for training increases. 

In this work, we apply recently proposed models of the scaling behavior of neural language models to SLMs, and use them to estimate the scale at which our current methods will match the linguistic performance of Large Language Models (LLMs), generative text-based systems that have achieved remarkably strong performance across a wide range of NLP applications \cite{brown2020language}. The main contributions of this work are:

\begin{figure*}[ht]
\vskip 0.2in
\begin{center}
\includegraphics[width=0.66666666666\columnwidth]{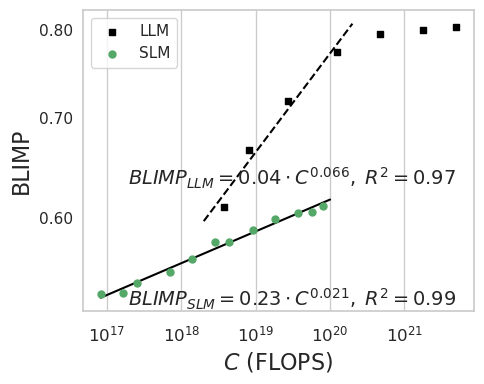}
\includegraphics[width=0.66666666666\columnwidth]{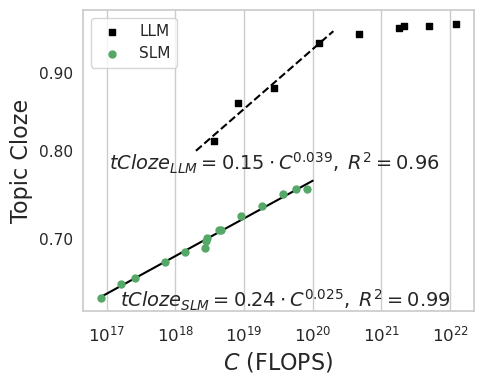}
\includegraphics[width=0.66666666666\columnwidth]{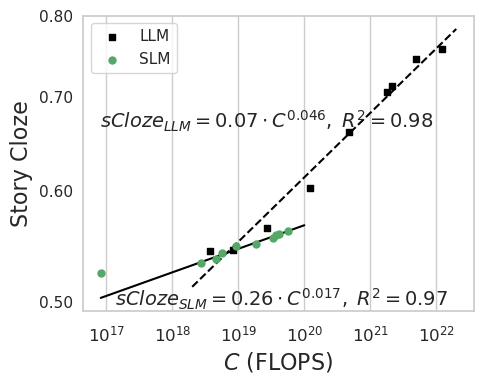}
\caption{Downstream linguistic performance scaling with compute for LLMs and SLMs. Axes are in logarithmic scale. Syntactic (BLIMP) and semantic (Topic Cloze and Story Cloze) metrics follow a power law before starting to saturate. Linguistic performance scales up to three orders of magnitude more slowly in SLMs relative to LLMs.
% The {\textcolor{red}{$\bigstar$}} shows when SLMs would match the performance of LLMs by extrapolating the power law.
}
\label{fig:downstream-scaling}
\end{center}
\vskip -0.2in
\end{figure*}

\begin{itemize}
    \item We trained over 50 SLMs with different number of parameters and data budgets. We show that the test loss of SLMs follows scaling power laws as those observed in text-based LLMs (Figure \ref{fig:test-scale}), and use the methods from \citet{hoffmann2022training} and \citet{muennighoff2023scaling} to model the scaling behavior of SLMs.
    \item We establish a strong correlation between the test loss of neural LMs and the downstream metrics commonly used to evaluate their syntactic and semantic abilities. Therefore, the linguistic performance of LMs follows similar scaling laws (Figure \ref{fig:downstream-scaling}). We leverage this insight to determine the relative efficiency with scale of SLMs relative to LLMs.
    
    \item We speculate that SLMs require more context than fits in their context window to acquire from commonly used speech datasets the semantic understanding measured by our metrics. Accordingly, we propose a new speech dataset to boost semantic understanding in SLMs. Specifically, we synthesized a spoken version of the Tiny Stories dataset \cite{eldan2023tinystories}, and show that its use during pre-training improves downstream semantic performance.
    \item On the basis of our previous observation, we studied the use of unigram tokenization to shorten sequences and pack more information in the context window of SLMs. However, our results suggest that a coarser tokenization is detrimental to downstream performance.
    
\end{itemize}

\section{Background}

\subsection{Generative spoken language modeling}
\label{sec:gslm}

We follow the GSLM framework from \citet{lakhotia-etal-2021-generative}. The general GSLM pipeline is composed of three separately trained models: (i) a speech tokenizer, (ii) a language model, and (iii) a vocoder (token-to-waveform) module. In the following, we provide background for the speech tokenizer and LM, as these are the components we use in this work. For details about the vocoder please refer to \citet{lakhotia-etal-2021-generative}.

\textbf{Speech tokenizers} transform raw speech waveforms into discrete representations. A speech encoder is used to extract continuous representations that are then transformed into discrete sequences through vector quantization. Formally, let $\mathcal{X} \in \mathbb{R}$ denote the domain of audio samples, a waveform is therefore a sequence of samples $x = (x_1, \dots, x_T)$, where $x_t \in \mathcal{X}$ for all $1 \leq t \leq T$. An encoder $F: \mathcal{X}^{m} \to \mathbb{R}^d$ transforms windows of samples of width $m$ into $d$ dimensional continuous frame representations. Applying $F$ to $x$ yields a sequence of frame representations $z = (z_1, \dots, z_{T'})$, where usually $T' < T$. Subsequently, a k-means algorithm is applied to the encoder output to generate a sequence of discrete speech tokens $u = (u_1, \dots, u_{T'})$, where $u_i \in \{1, \dots, K\}$ for $1 \leq i \leq T'$, and $K$ is the size of the vocabulary.

\textbf{Language models} aim to learn the joint probability of token sequences $P(w_1, \dots, w_n)$. By the chain rule of probability, the probability of a sequence can be computed as a product of its conditional probabilities:

\begin{equation}
    P(w_1, \ldots, w_n) = \prod_{i=1}^{n} P(w_i | w_1, \ldots, w_{i-1})
    \label{eq:lm}
\end{equation}

Neural LMs, parameterized by $\theta$, are neural networks that model the conditional probabilities $P_{\theta}(w_i | M(w_1, \ldots, w_{i-1}))$, where $M$ is a representation of the previous tokens. The network is optimized to minimize the negative log-likelihood of observed ground truth sequences:

\begin{equation}
    L = -\sum_{i=1}^n P_{\theta}(w_i | M(w_1, \ldots, w_{i-1}))
    \label{eq:lm_loss}
\end{equation}

Nowadays, the network is typically a transformer \cite{vaswani17transformer}. LLMs are large transformer LMs trained on large text corpora (billions of parameters and tokens). SLMs are neural LMs applied to speech tokens $u$. 

\subsection{Scaling laws for neural language models}
\label{sec:scaling_laws}

The performance of deep learning models often behaves predictably as a function of model size, dataset size, and compute \cite{hestness2017}. \citet{kaplan2020} showed that the loss $L$ (Equation \ref{eq:lm_loss}) of large neural LMs scales with a power law behavior as a function of these three scale factors:
\begin{equation}
\begin{aligned}
L(C) \propto C^{\gamma}, \quad L(N) \propto N^{\alpha}, \quad L(D) \propto D^{\beta}
\end{aligned}
\end{equation} Where $C$ is the amount of compute (in FLOPS), $N$ is the number of parameters of the model, and $D$ is the number of training tokens.

Building upon their work, \citet{hoffmann2022training} proposed a parametric function to model the final loss of neural LMs trained for a single epoch as a function of $N$ and $D$:

\begin{equation}
    \hat{L}(N, D) = E + \frac{A}{N^{\alpha}} + \frac{B}{D^{\beta}}
    \label{eq:chinchilla}
\end{equation} Where the first term is the loss for an ideal LM, and should correspond to the entropy of the distribution of token sequences. The second term captures the approximation error that results from using a neural network with $N$ parameters to approximate the ideal generative process. The final term reflects that the model is not trained to convergence, as a finite number of optimization steps are performed on a sample of size $D$ from the real distribution.

\citet{hoffmann2022training} aimed to solve the problem of optimal allocation of resources given a fixed compute budget $C_{\text{avail}}$. They proposed to approximate the compute needed to train a transformer LM with $N$ parameters on $D$ tokens as $C\approx6ND$. Then, the problem of optimal allocation of compute for model size and training data is:

\begin{equation}
\min_{N, D} \hat{L}(N, D), \quad \text{s.t.} \quad 6ND = C_{\text{avail}}
\label{eq:chinchilla_opt}
\end{equation} For which the solution is:

\begin{equation}
\begin{aligned}
N_\text{opt}(C) &= G \left(\frac{C}{6}\right)^a \\ \, D_\text{opt}(C) &= \frac{1}{G} \left(\frac{C}{6}\right)^b
\end{aligned}
\label{eq:chinchilla_sol}
\end{equation} With: 

\begin{equation*}
    G = \left(\frac{\alpha A}{\beta B}\right)^{\frac{1}{\alpha + \beta}},\, a = \frac{\beta}{\alpha + \beta},\text{ and }b = \frac{\alpha}{\alpha + \beta}
\end{equation*}

\citet{muennighoff2023scaling} generalized Equation \ref{eq:chinchilla} to the case of multi-epoch training by replacing $D$ and $N$ with terms corresponding to the effective data $D'$ and effective model parameters $N'$:

\begin{equation}
    \hat{L}(N', D') = E + \frac{A}{N'^{\alpha}} + \frac{B}{D'^{\beta}}
    \label{eq:chinchilla_multiep}
\end{equation} Where $D' \leq D$ is the number of effective training tokens, assuming that the value of repeated tokens decays exponentially. Similarly, they note that oversized models offer diminishing returns per parameter, as excess parameters learn the same features and do not add value (in the extreme). They propose an exponential decay model for them, yielding a number of effective parameters $N' \leq N$. They derived the expressions for $D'$ and $N'$ as:

\begin{equation}
    \begin{aligned}
        D' &= U_D + U_D R_D^* (1 - e^{\frac{-R_D}{R_D^*}})\\
        N' &= U_N + U_N R_N^* (1 - e^{\frac{-R_N}{R_N^*}})
    \end{aligned}
\end{equation} Where $U_D$ is the number of unique tokens used, $R_D = \frac{D}{U_D} - 1$ is the number of repetitions (0 for a single epoch), $U_N$ is the number of parameters needed to optimally fit $U_D$ according to Equation \ref{eq:chinchilla_sol}, $R_N = \frac{N}{U_N} - 1$ is the number of excess parameters, and $R_D^*$ and $R_N^*$ are constants.

The constants $E$, $A$, $B$, $\alpha$, $\beta$, $R_D^*$ and $R_N^*$ can be estimated empirically by fitting Equation \ref{eq:chinchilla} or \ref{eq:chinchilla_multiep} to a set of tuples $(N, D, R_N, R_D, L)$ obtained from training experiments with different budgets.
 
% Given a set of neural LM training runs yielding a set of $(L, N, D, )$ tuples, we can empirically estimate the constants $E$, $A$, $B$, $\alpha$ and $\beta$ by minimizing the error between the predicted loss and observed loss:

% \begin{equation}
%     \min_{E, A, B, \alpha, \beta} \sum_{\text{Runs}\;i} G(\hat{L}(N_i, D_i) - L_i)  
% \end{equation} Where $G$ is some error function.

% Equation \ref{eq:chinchilla} can be used to determine the optimal $N$ and $D$ to minimize $L$ given a compute budget $C$. In transformer LMs, $C \approx 6ND$. 

\section{Experimental setup}

% \subsection{Setup}

\subsection{Models and training}
\label{sec:models}

We adhere to the framework described in Section \ref{sec:gslm}. For the speech tokenizer, we use a pre-trained HuBERT model \cite{hsu2021hubert} with frame-rate of 25 Hz as the speech encoder $F$, and a vocabulary size of $K = 500$. This setup reports the best performance among publicly available models \cite{hassid2023textually}. For the SLMs we use the Llama architecture \cite{touvron2023llama} with context window of 2050 tokens. Table \ref{tab:models} describes the model sizes used in our experiments. For the LLMs, we use the Pythia suite of pre-trained LLMs \cite{biderman23}, ranging in size from 14M to  6.9B parameters (we do not use the largest 12B model), and trained with $\sim$300B tokens.

All SLMs are optimized using AdamW \cite{loshchilov2018decoupled} with weight decay of 0.1, maximum learning rate of 5e-4, half-cycle cosine decay learning rate schedule to 5e-5, and a warm-up initial stage of $\max(100, 0.01 \, n_{iters})$ steps, where $n_{iters}$ is the number of training steps, which varies for each experiment according to the data budget. We use batch sizes of 64, 128, 256 and 512 for the models with 20M, 85M, 155M and 309M, and 828M parameters, respectively.

\begin{table}[t]
% \vskip 0.15in
\begin{center}
\begin{small}
\setlength\extrarowheight{-3pt}
\begin{sc}
\begin{tabular}{lcccr}
\toprule
Size & Layers & Model dim. & Heads \\
\midrule
20M & 6 & 512 & 8 \\
85M & 12 & 768 & 12 \\
155M & 12 & 1024 & 16 \\
309M & 24 & 1024 & 16 \\
823M & 16 & 2048 & 32 \\
\bottomrule
\end{tabular}
\end{sc}
\end{small}
\end{center}
\caption{Models description.}
\label{tab:models}
% \vskip -0.1in
\end{table}

To fit the constants in Equations \ref{eq:chinchilla} and \ref{eq:chinchilla_multiep}, we adopt the approaches of \citet{hoffmann2022training} and \citet{muennighoff2023scaling}, utilizing the Huber loss with $\delta=0.03$ as the error function and L-BFGS as optimizer. Following \citet{muennighoff2023scaling}, we first fit the parameters $E$, $A$, $B$, $\alpha$, and $\beta$ using the single-epoch runs, and afterwards fit $R^*_D$ and $R^*_N$ using the multi-epoch runs.

\subsection{Evaluation}

We use the \textsc{sBLIMP} task \cite{sblimp} to measure syntactic performance. In \textsc{sBLIMP}, the model is presented with a matched pair of speech segments, grammatical and ungrammatical sentences. The objective is to assign higher probability to the grammatical sentence. 

To evaluate semantic understanding we use the spoken \textsc{StoryCloze} benchmark from \citet{hassid2023textually}, a spoken version of the StoryCloze textual benchmark \cite{mostafazadeh-etal-2016-corpus}, which consists of 4k five-sentence commonsense stories. In StoryCloze, the model receives as input the first four sentences of a story, and has to assign higher probability to the correct final sentence than to an adversarial negative sample. The spoken benchmark comes in two versions: Story Cloze and Topic Cloze. The difference between them lies in how the negative sample is generated. Spoken Story Cloze uses the same samples as the textual benchmark, which require commonsense reasoning to distinguish from the real ending. In Topic Cloze, the negatives are randomly sampled from the whole dataset, and therefore measures the ability of the model to stay on topic.

Regarding upstream performance, in all cases we report and use for the parametric fits the average loss (Equation \ref{eq:lm_loss}) on the test set.

\subsection{Data}
\label{sec:data}

\begin{table}[t]
\begin{center}
\setlength\extrarowheight{-3pt}
\begin{small}
\begin{sc}
\begin{tabular}{lcccr}
\toprule
Dataset & Hours & \makecell{HuBERT \\ Tokens} & Unigram \\
\midrule
LibriSpeech    & 960 & 67M & 38M\\
LibriLight & 53k & 3.74B & 2.11B \\
SWC    & 1k & 32M & 19M \\
Tedlium    & 1.6k & 0.11B & 67M \\
People     & 7k & 0.48B & 0.29B \\
Vox Populi      & 24k & 1.64B & 1.08B \\
sTinyStories     & 72k & 4.82B & 2.71B \\
\midrule
Total & 160k & 10.89B & 6.31B\\
\bottomrule
\end{tabular}
\end{sc}
\end{small}
\end{center}
\caption{Datasets statistics. The \textsc{Unigram} column corresponds to the dataset of HuBERT tokens compressed through unigram tokenization.}
\label{tab:data}
\end{table}

\subsubsection{Datasets}

We use a collection of publicly available English speech datasets for training: LibriSpeech \cite{librispeech}, LibriLight \cite{librilight}, SWC \cite{swc}, Tedlium \cite{tedlium}, People's Speech \cite{people}, and Vox Populi \cite{voxpopuli}; and a novel dataset: \textsc{sTinyStories}, a spoken version of the Tiny Stories dataset \cite{eldan2023tinystories} that we synthesized using the single-speaker TTS system provided by \citet{wang-etal-2021-fairseq}. Tiny Stories is a synthetic text corpus of short stories designed to boost commonsense reasoning in neural LMs. We propose \textsc{sTinyStories} because we hypothesize that the semantic understanding that tasks such as Story Cloze measure is hard to acquire from commonly used speech datasets. Consider for instance the audiobooks in LibriLight. The data has long-range dependencies spanning multiple pages, whereas our SLMs can ingest roughly a dozen sentences of spoken text in their context window. Other datasets, which were mainly designed to serve as training data for automatic speech recognition systems, consist of too small fragments of audio that lack meaningful causal structure. \textsc{sTinyStories} consists of full stories with causal structure that fit within the context window of our SLMs.

We do not include samples from \textsc{sTinyStories} in our test set, as we intend to use our test loss as measure of the quality with which SLMs model natural language, not synthetic one.  For other datasets we use the defined held-out sets for testing. In cases where a held-out set is not defined, we randomly sampled 1\% of the data to serve as test set. See Table \ref{tab:data} for dataset sizes. 

\subsubsection{Data budgets}
\label{sec:data_budgets}

In order to have a representative set of samples to fit Equations \ref{eq:chinchilla} and \ref{eq:chinchilla_multiep}, for each model size, we performed training runs with a ratio of training tokens $D$ to parameters $N$: $D / N \in \{2, 4, 8, 10, 20, 32, 64, 100\}$. This setup yields single-epoch and multi-epoch runs for the larger models but not for the smaller models (e.g. for the model with 85M parameters the maximum number of training tokens corresponds to 0.99 epochs). To  better fit Equation \ref{eq:chinchilla_multiep}, we performed additional experiments so that for each model size there were runs with training epochs in $\{2, 4, 8, 10\}$, with the exception of the 828M parameter model, for which the maximum was 8 epochs.

\section{Results}

\subsection{Gains from sTinyStories}
\label{sec:stiny}

\begin{figure}[t]
\vskip 0.2in
\begin{center}
\includegraphics[width=0.7\columnwidth]{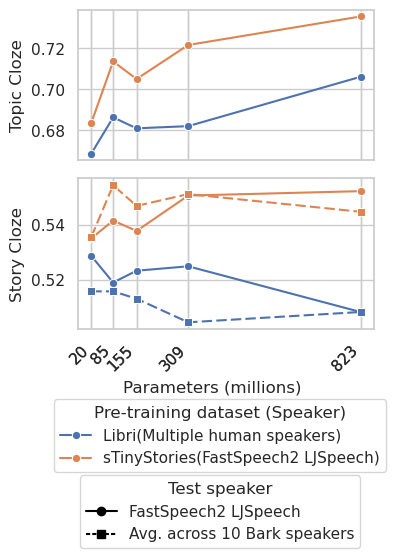}
\caption{Gains from synthetic data on downstream semantic performance of SLMs. Pre-training on sTinyStories yields consistent improvements on semantic understanding relative to pre-training on audiobooks (LibriSpeech plus LibriLight). Performance gains hold for mismatched train and test speakers.}
\label{fig:stiny}
\end{center}
\vskip -0.2in
\end{figure}

In order to determine if \textsc{sTinyStories} meaningfully contributes to the semantic understanding of SLMs, we compare the performance on Topic Cloze and Story Cloze of models trained on one epoch of the union of LibriSpeech and LibriLight, against models trained on an equivalent amount of \textsc{sTinyStories} tokens. Figure \ref{fig:stiny} shows the obtained results. Models trained on \textsc{sTinyStories} consistently outperform those trained on audiobooks across all model scales. A factor that could contribute to the observed performance gain is the match between training and evaluation speakers, as both \textsc{sTinyStories} and Story Cloze were synthesized using the single-sepaker TTS from \citet{wang-etal-2021-fairseq}. However, we believe this to be unlikely as the speech tokenizer we use likely captures little speaker-specific information \cite{nguyen23_interspeech}. To isolate the potential impact of speaker mismatch between training and evaluation data, we created a multi-speaker version of the Story Cloze benchmark using Bark TTS \footnote{https://github.com/suno-ai/bark}, and repeat the evaluations. The results, also shown in Figure \ref{fig:stiny}, indicate that even with mismatched train and test speakers training on \textsc{sTinyStories} yields performance gains.

\subsection{Benchmarking our setup}

To validate our setup, we compared our best performing model with other models in the SLM literature in Table \ref{tab:benchmark}. Our model outperformed all other speech-only LMs on the semantic tasks, and performed second best in general, even relative to hybrid speech-text LMs. Notably, our model outperformed models with a larger compute budget. Considering that the models from \citet{hassid2023textually} and \citet{nguyen2024spiritlm} use similar hyperparameters (same speech tokenizer and the Llama architecture for LMs); the most likely factor to explain the performance difference is the data used. We believe these results further illustrate the benefits from using \textsc{sTinyStories}.

\begin{table*}[t]
% \vskip 0.15in
\begin{center}
\begin{small}
\begin{sc}
\begin{tabular}{lccccc}
\hline
                                            & Parameters & Tokens               & BLIMP               & Topic Cloze          & Story Cloze          \\ \hline
\multicolumn{6}{l}{\textit{Speech-only language models}}                                                                                                      \\
GSLM \cite{lakhotia-etal-2021-generative}       & ~100M      & -                    & 54.2                 & 66.6                 & 53.3                 \\
AudioLM \cite{borsos-audiolm}                      & ~150M      & -                    & \underline{\textbf{64.7}}        & -                    & -                    \\
\citet{hassid2023textually}, Cold-init 1.3B & ~1.3B      & ~10.8B               & 56.5                 & -                    & -                    \\
\citet{nguyen2024spiritlm}                        & ~7B        & ~100B                & 58.0                 & 72.9        & 54.8        \\
Ours (best model)                                  & 823M       & ~82B                 & \textbf{61.3}     & \textbf{78.0}     & \textbf{56.7}     \\ \hline
\multicolumn{6}{l}{\textit{Speech language models initialized from text language models}}                                                                     \\
TWIST \cite{hassid2023textually}                 &            & \multicolumn{1}{l}{} & \multicolumn{1}{l}{} & \multicolumn{1}{l}{} & \multicolumn{1}{l}{} \\
$\quad$- Warm-init 1.3B                            & ~1.3B      & ~10.8B               & 57.1                 & 70.6                 & 52.4                 \\
$\quad$- Warm-init 7B                              & ~7B        & ~36B                 & 59.0                 & 74.1                 & 55.1                 \\
$\quad$- Warm-init 13B                             & ~13B       & ~36B                 & 59.2                 & 76.4                 & 55.4                 \\ \hline
\multicolumn{6}{l}{\textit{Mutltimodal speech-text language models initialized from text language models}}                                                    \\
SpiRit-LM \cite{nguyen2024spiritlm}                        & ~7B        & ~100B                & 58.3                 & \underline{\textbf{82.9}}        & \underline{\textbf{61.0}}        \\ \hline
\multicolumn{6}{l}{\textit{Toplines}}                                                                                                             \\
Pythia \cite{biderman23} 6.9B                                 & ~6.9B      & ~300B                & 80.0                 & 97.5                 & 76.21                \\
Human \cite{hassid2023textually}                                      & -          & -                    & -                    & 90.2                 & 79.9                 \\ \hline
\end{tabular}
\end{sc}
\end{small}
\end{center}
\caption{Models benchmarking. The best model resulting from our experiments obtains the best semantic performance across speech-only models, and the second best overall in all tasks.}
\label{tab:benchmark}
\end{table*}

\subsection{Scaling laws}

We trained multiple SLMs for each model size with different data budgets as described in Section \ref{sec:data_budgets}. The resulting learning curves for single-epoch runs are presented in Figure \ref{fig:test-scale} as a function of compute, and show that the envelope of minimal loss per FLOP follows a power law.

\subsubsection{Downstream scaling with compute}

\begin{figure*}[ht]
\vskip 0.2in
\begin{center}
\includegraphics[width=0.6\columnwidth]{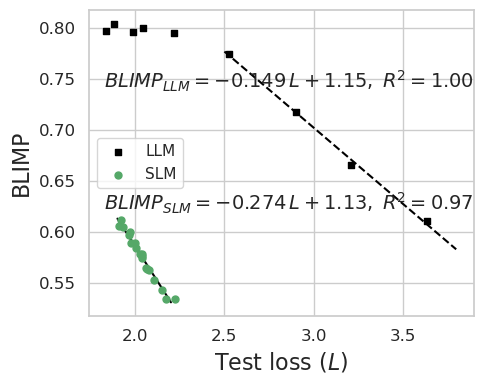}
\includegraphics[width=0.6\columnwidth]{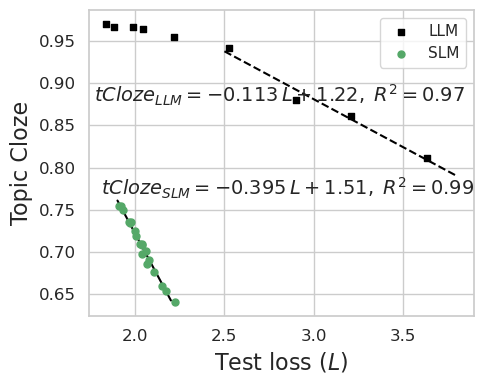}
\includegraphics[width=0.6\columnwidth]{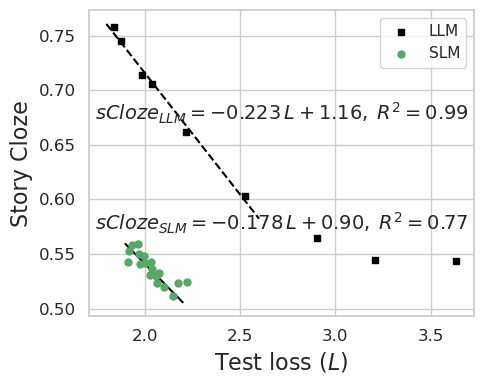}
\caption{Correlation between downstream linguistic performance and test loss for LLMs and SLMs. Syntactic (BLIMP) and semantic (Topic Cloze and Story Cloze) metrics are strongly linearly correlated with the upstream test loss before saturation.
% The {\textcolor{red}{$\bigstar$}} shows when SLMs would match the performance of LLMs by extrapolating the power law.
}
\label{fig:downstream-corr}
\end{center}
\vskip -0.2in
\end{figure*}

We analyzed the relationship between the upstream and linguistic downstream performance in SLMs and LLMs. Figure \ref{fig:downstream-corr} shows the obtained results. Downstream linguistic metrics before saturation are strongly correlated with the upstream test loss in both LLMs and SLMs. Therefore, the envelope of maximum downstream performance per FLOP also follows a power law, i.e. for a downstream performance function $Q$, $Q \propto C^{\gamma_q}$. The power laws for the different performance metrics are presented in Figure \ref{fig:downstream-scaling} and the exponents in Table \ref{tab:llmvsslm}.

These results allow us to compare the efficiency with scale of LLMs and SLMs. For each metric, we can interpret the ratio between the $\gamma_q$ exponents of the power laws of LLMs and SLMs as the relative efficiency with scale. For BLIMP, the ratio is $\frac{0.066}{0.021} = 3.14$, indicating that for an increase in compute $\Delta C$ yielding a $\Delta Q$ in LLM's syntactic performance, SLMs require $10^{3.14} \Delta C$ to get the same $\Delta Q$. Similarly, for Topic Cloze and Story Cloze the ratios are $1.56$ and $2.7$, respectively.

\begin{table}[t]
\begin{center}
\begin{small}
\begin{sc}
\setlength\extrarowheight{-3pt}
\begin{tabular}{@{}lccc@{}}
\toprule
\multicolumn{1}{c}{\multirow{2}{*}{Modality}} & \multicolumn{3}{c}{$\gamma_q$} \\ \cmidrule(l){2-4} 
\multicolumn{1}{c}{}                          & BLIMP   & tCloze   & sCloze  \\ \midrule
Text                                           & 0.066   & 0.039    & 0.046   \\
Speech                                           & 0.021   & 0.025    & 0.017   \\ \bottomrule
\end{tabular}
\end{sc}
\end{small}
\end{center}
\caption{$\gamma_q$ power law coefficients of downstream performance with compute as depicted in Figure \ref{fig:downstream-scaling}.}
\label{tab:llmvsslm}
\end{table}

\subsubsection{Scaling with parameters and tokens}

We fitted the functions from Equations \ref{eq:chinchilla} and \ref{eq:chinchilla_multiep} to our data using the procedure described in Section \ref{sec:models}. We present the empirically fitted scaling law parameters and compare them to the ones obtained for text by \citet{muennighoff2023scaling} in Table \ref{tab:chinchilla-params}. From Equation \ref{eq:chinchilla_sol}, $N_{opt} \propto C^a$ and $D_{opt} \propto C^b$. For both modalities $a \approx b \approx 0.5$, suggesting that as compute increases, model size and data should be scaled equally for optimal performance. Contrary to text, $R^*_N > R^*_D$, indicating that repeated tokens decay faster than excess parameters (albeit both slower than in text). Therefore, in SLMs, compute allocated to parameters should scale faster than compute allocated for epochs. 

\begin{table}[t]
\begin{center}
\begin{footnotesize}
\setlength\extrarowheight{-4pt}
\begin{sc}
% \begin{tabular}{cccccc}
% \toprule
%  & E & A & B & $\alpha$ & $\beta$ \\
% \midrule
% Text & 1.69 & 406.4 & 410.7 & 0.34 & 0.28 \\
% \midrule
% % \makecell{Speech \\ (25 Hz, \\ $K=500$)} & 1.73 & 13.92 & 39.80 & 0.25 & 0.24 \\\midrule
% \makecell{Speech} & 1.73 & 13.92 & 39.80 & 0.25 & 0.24 \\
% \midrule
% \makecell{Speech \\ (Unigram)} & 1.42 & 3.85 & 8.90 & 0.15 & 0.16 \\
% % Unigram & 53k & 3.74B & 2.11B
% \bottomrule
% \end{tabular}
\setlength{\tabcolsep}{2.25pt}
\begin{tabular}{@{}cccccccc@{}}
\toprule
                               & E    & A     & B     & $\alpha$ & $\beta$ & $R^*_N$ & $R^*_D$ \\ \midrule
% \makecell{Text \\ \scriptsize{\citeauthor{hoffmann2022training}}}                           & 1.69 & 406 & 411 & 0.34     & 0.28    & -    & -   \\ \midrule
\makecell{Text \\ \scriptsize{\citeauthor{muennighoff2023scaling}}}                           & 1.87 & 521 & 1488   & 0.35   & 0.35    & 5.31 & 15.4  \\ \midrule
\makecell{Speech}              & 1.73 & 13.9 & 39.8 & 0.25     & 0.24    & 31.0   & 25.0   \\ \midrule
\makecell{Speech \\ (Unigram)} & 1.42 & 3.85  & 8.90  & 0.15     & 0.16    & -       & -       \\ \bottomrule
\end{tabular}
\end{sc}
\end{footnotesize}
\end{center}
\caption{Scaling law parameters fit to Equations \ref{eq:chinchilla} and \ref{eq:chinchilla_multiep} for different language tokenizations.}
\label{tab:chinchilla-params}
\end{table}

% \begin{table}[t]
% \begin{center}
% \begin{small}
% \begin{sc}
% \begin{tabular}{@{}ccc@{}}
% \toprule
%                                & $R^*_N$ & $R^*_D$ \\ \midrule
% Text                           & 5.31   & 15.39    \\
% \makecell{Speech \\ (Unigram)} & 31.02   & 24.94   \\ \bottomrule
% \end{tabular}
% \end{sc}
% \end{small}
% \end{center}
% \caption{Scaling law parameters fit to Equation \ref{eq:chinchilla_multiep} for different language tokenizations.}
% \label{tab:chinchilla-params-multiep}
% \end{table}

\begin{figure*}[ht]
\vskip 0.2in
\begin{center}
\includegraphics[width=0.5\columnwidth]{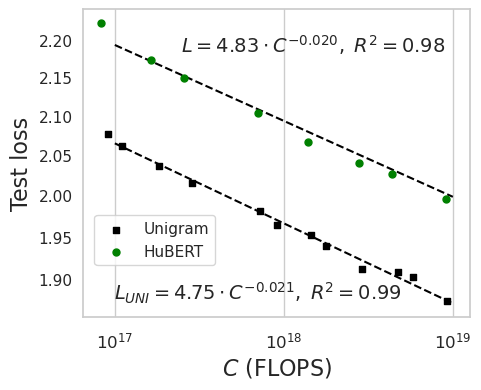}
\includegraphics[width=0.5\columnwidth]{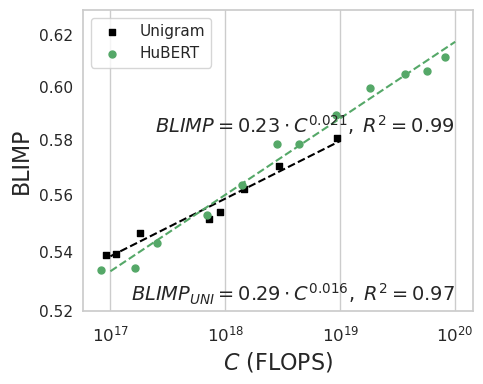}
\includegraphics[width=0.5\columnwidth]{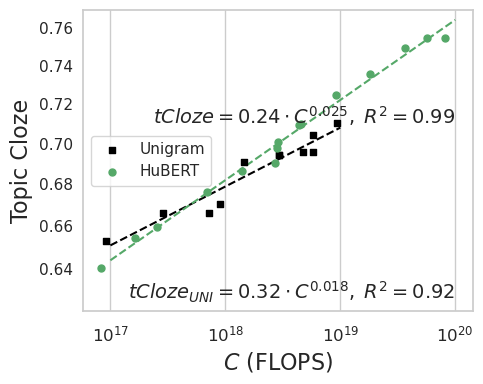}
\includegraphics[width=0.5\columnwidth]{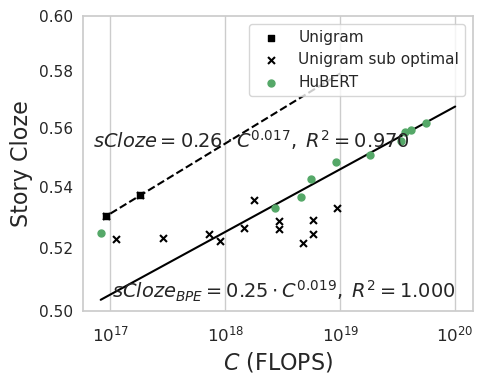}
\caption{Comparison of the scaling behavior of SLMs trained on raw speech tokens and unigram compressed tokens. Axes are in logarithmic scale. The upstream loss of SLMs trained on unigram tokens scales better with compute, but downstream performance scales worse. Notably, the Story Cloze metric for SLMs trained on unigram tokens does not seem to improve with increased compute.
% The {\textcolor{red}{$\bigstar$}} shows when SLMs would match the performance of LLMs by extrapolating the power law.
}
\label{fig:unigram-scaling}
\end{center}
\vskip -0.2in
\end{figure*}

\subsection{Unigram tokenization}

As mentioned in Section \ref{sec:data}, we believe that the limited context window of SLMs could cripple their ability to model the long-range dependencies in language required for causal reasoning. Seeking to mitigate this limitation,  we apply unigram tokenization to shorten the length of speech token sequences. We use the SentencePiece tokenizer \cite{kudo-richardson-2018-sentencepiece} with a vocabulary size of 5000. We choose the vocabulary size on the scale of previous works that have used similar tokenization strategies for speech applications \cite{chang2023exploring}. The resulting dataset sizes after compression are presented in Table \ref{tab:data}.

We train a set of Speech LMs on the compressed datasets, with model sizes up to 309M parameters and data budgets ranging from 740M to 6.31B tokens. We analyze the scaling behavior of the upstream and downstream metrics and compare it with SLMs trained on raw HuBERT speech tokens in Figure \ref{fig:unigram-scaling}. SLMs trained on unigram compressed speech tokens show similar upstream scaling with compute, but worse downstream scaling. Notably, the performance on the StoryCloze benchmark does not seem to scale with compute. 

We fitted the function from Equation \ref{eq:chinchilla} to the results obtained on the compressed dataset. Table \ref{tab:chinchilla-params} presents the resulting scaling law parameters. Similar to the previous findings, for a given compute budget, scaling model size and training data equally is optimal for performance. Due to the poor downstream results obtained with unigram tokenization and the lack of sufficient compute resources, we did not perform multi-epoch training experiments.

\section{Related work}

Previous works have studied the scaling behavior of neural networks on speech applications. \citet{droppo2021scaling} showed that acoustic models trained with an auto-predictive coding loss follow similar power laws to those observed in neural LMs. \citet{Aghajanyan23} used the scaling laws from \citet{hoffmann2022training} to model the scaling behavior of the upstream loss of neural LMs on multiple modalities, including speech. They used a speech tokenizer with higher framerate (50 Hz) and vocabulary size ($K = 2000$) than the one we used (Section \ref{sec:models}). Such fine-grained tokenizers capture a lot of the paralinguistic information in speech \cite{nguyen23_interspeech}. Therefore, their speech tokens can be considered almost a different modality due to the acoustic variance. Furthermore, they do not study the behavior with scale of downstream performance. In this work, we focus on the linguistic content of the signal. As reported by \citet{hassid2023textually}, our speech tokenizer performs best on downstream linguistic applications, and is therefore a more suitable choice to study the scaling behavior of the linguistic performance of SLMs.

This paper is most closely related to the work of \citet{hassid2023textually}. We largely follow their setup in terms of hyperparameters and evaluation metrics. They reported improved linguistic downstream performance with scale in SLMs, but did not characterize their scaling behavior. Our scaling laws allow practitioners to determine the compute needed to attain a specific loss, syntactic and/or semantic performance; and its optimal allocation with respect to parameters and tokens. To the best of our knowledge, we are the first to model the scaling properties of downstream linguistic performance in SLMs, and to study the scaling of the considered downstream metrics on text-based LLMs. This enables a comparison between the two modalities in terms of scaling efficiency.

\section{Discussion}

% \subsection{On our results}

Our work showed that the upstream and downstream linguistic performance of our current methods for GSLM scales predictably with compute. This suggests that with sufficient computational resources, the goal of the textless NLP project of achieving neural LMs trained exclusively on speech, and matching the linguistic proficiency of their text-based counterparts, is achievable. The cost of such models could be prohibitive though, as we estimate that they will require up to three orders of magnitude more compute than a text-based LLM to achieve equivalent performance. In this regard, recent methods that leverage transfer learning from text-based LLMs \cite{hassid2023textually, zhang-etal-2023-speechgpt, nguyen2024spiritlm} are likely to be a better choice to achieve highly performant generative speech models. These hybrid text-speech generative models often enable cross-modal applications such as ASR or TTS. However, it remains to be seen how knowledge transfer from LLMs performs when the speech data is in a different language than the one the LLM was trained on. If there is no significant cross-lingual knowledge transfer between text and speech modalities, SLMs could still be an attractive choice for low-resource languages.

We explored the use of synthetic data and coarser tokenization to increase the semantic abilities of SLMs. Our synthetic dataset improved semantic performance, but using a coarser tokenization led to overall degradation of downstream performance. We do not have yet an hypothesis for why coarser tokens degrade performance, as this seems counter-intuitive, and contradicts the findings on other speech applications \cite{chang2023exploring}. We leave this as an interesting issue to address in future work. Moreover, we believe that working on methods that allow to increase the information density per context-window of SLMs holds promise to improve their scaling behavior.

\section{Limitations}

Any extrapolation from our models of the scaling behavior of SLMs should be considered optimistic for the following reasons: \textbf{1)} Our models for downstream performance ignore the fact that the metrics saturate. As observed in text LLMs, the improvements with scale slow down as performance approaches the saturation value. It is likely that, due to saturation, the compute required to yield a particular performance will be larger than predicted. Moreover, due to the lower density of linguistic information per context window in SLMs relative to LLMs, the saturation values of the metrics may be lower for SLMs. \textbf{2)} The LLMs from the Pythia suite that we used in this study are likely overtrained (all models were trained with $\sim$300B tokens). Optimally trained LLMs (according to Equation \ref{eq:chinchilla_sol}) should show better performance with scale, and therefore widen the gap with the scaling efficiency of SLMs. \textbf{3)} The envelope of minimal loss per FLOP (Figure \ref{fig:test-scale}) might show a slight negative curvature at larger scale \cite{hoffmann2022training}, reducing the scaling efficiency.

% \begin{enumerate}
% \item 

% \item

% \item

% \end{enumerate}

% It is possible that at larger compute budgets there will be deviations from the observed scaling behavior. For instance, in text LLMs trained with more compute than that considered in this work, the envelope of minimal loss per FLOP shows a slight negative curvature \cite{hoffmann2022training}. Depending on whether this phenomenon would also occur in SLMs, and to what degree, the observed relative efficiency could change, although likely not significantly. 

% Our models for downstream performance assume that the scaling behavior will hold at larger scale and ignore the fact that the metrics saturate. As observed in text LLMs, the improvements with scale slow down as performance approaches the saturation value. Therefore, predictions made using our models for larger compute budgets should be considered optimistic, as it is likely that due to saturation the compute required to yield a particular performance will be larger than predicted.  

% Finally, it should be noted that the text LLMs from the Pythia suite we used are likely overtrained (all models were trained with $\sim$300B tokens). Therefore, it is likely that optimally-trained (according to Equation \ref{eq:chinchilla_opt}) text LLMs would show even better performance with scale, and therefore widen the gap with the scaling efficiency of SLMs.
% If we are to expect a similar phenomena on speech, then our estimates should be considered optimistic. 

\section{Conclusions}

We have trained a large set of SLMs with different compute budgets and studied the scaling properties of their upstream and downstream performance using recently proposed models of scaling laws for neural LMs. The obtained models allow practitioners to optimally allocate compute to attain a specific loss, syntactic, and/or semantic performance. We showed that the pre-training loss and downstream linguistic performance of SLMs and LLMs is highly correlated, and both scale predictably according to power laws. This allowed us to compare the scaling properties of SLMs and LLMs, from which we established that the linguistic abilities of SLMs scale up to three orders of magnitude more slowly. Additionally, we proposed a new speech dataset, \textsc{sTinyStories}, and showed that its use during pre-training improves downstream semantic performance. Finally, we explored the use of coarser speech tokenization as a method to increase the amount of tokens per context window in SLMs, but obtained worse downstream performance.

\section*{Acknowledgements}

We are grateful to the French National Research Agency for their support through the ANR-20-CE23-0012-01 (MIM) grant, and the Institute of Convergence ILCB, supported by grants from France 2030 (ANR-16-CONV-0002) and the Excellence Initiative of Aix-Marseille University (A*MIDEX). This work was granted access to the HPC resources of GENCI-IDRIS under the allocation AD011014044.

% This document has been adapted
% by Steven Bethard, Ryan Cotterell and Rui Yan
% from the instructions for earlier ACL and NAACL proceedings, including those for 
% ACL 2019 by Douwe Kiela and Ivan Vuli\'{c},
% NAACL 2019 by Stephanie Lukin and Alla Roskovskaya, 
% ACL 2018 by Shay Cohen, Kevin Gimpel, and Wei Lu, 
% NAACL 2018 by Margaret Mitchell and Stephanie Lukin,
% Bib\TeX{} suggestions for (NA)ACL 2017/2018 from Jason Eisner,
% ACL 2017 by Dan Gildea and Min-Yen Kan, 
% NAACL 2017 by Margaret Mitchell, 
% ACL 2012 by Maggie Li and Michael White, 
% ACL 2010 by Jing-Shin Chang and Philipp Koehn, 
% ACL 2008 by Johanna D. Moore, Simone Teufel, James Allan, and Sadaoki Furui, 
% ACL 2005 by Hwee Tou Ng and Kemal Oflazer, 
% ACL 2002 by Eugene Charniak and Dekang Lin, 
% and earlier ACL and EACL formats written by several people, including
% John Chen, Henry S. Thompson and Donald Walker.
% Additional elements were taken from the formatting instructions of the \emph{International Joint Conference on Artificial Intelligence} and the \emph{Conference on Computer Vision and Pattern Recognition}.

\bibliography{custom}
\bibliographystyle{acl_natbib}

% \appendix

% \section{Bench}
% \label{sec:appendix_benchmark}

% This is an appendix.

\end{document}